# 3D PRINTED BRAIN-CONTROLLED ROBOT-ARM PROSTHETIC VIA EMBEDDED DEEP LEARNING FROM sEMG SENSORS

DAVID LONSDALE[1], LI ZHANG[1], RICHARD JIANG[2]

[1]Computer & Info Sciences, Northumbria University, Newcastle Upon Tyne, UK
[2]Computing & Communication, Lancaster University, Bailrigg, Lancaster, UK
E-MAIL: li.zhang@northumbria.ac.uk

**Abstract:**
In this paper, we present our work on developing robot arm prosthetic via deep learning. Our work proposes to use transfer learning techniques applied to the Google Inception model to retrain the final layer for surface electromyography (sEMG) classification. Data have been collected using the Thalmic Labs Myo Armband and used to generate graph images comprised of 8 subplots per image containing sEMG data captured from 40 data points per sensor, corresponding to the array of 8 sEMG sensors in the armband. Data captured were then classified into four categories (Fist, Thumbs Up, Open Hand, Rest) via using a deep learning model, Inception-v3, with transfer learning to train the model for accurate prediction of each on real-time input of new data. This trained model was then downloaded to the ARM processor based embedding system to enable the brain-controlled robot-arm prosthetic manufactured from our 3D printer. Testing of the functionality of the method, a robotic arm was produced using a 3D printer and off-the-shelf hardware to control it. SSH communication protocols are employed to execute python files hosted on an embedded Raspberry Pi with ARM processors to trigger movement on the robot arm of the predicted gesture.

**Keywords:**
Prosthetics, Brain Interface, Deep Learning on ARM.

## 1. Introduction

Prosthetic limbs have long been developed and can be sourced in their infant form as early as 1500 B.C. [1]. Although such developments were primitive in nature in the form of 'peg legs' and 'hand hooks', since the turn of the century, prosthetics have become highly sophisticated, computerised medical devices [2]. Performance and functionality have dramatically improved thanks to the extensive evolution of technology and manufacturing in domains such as smaller, faster microprocessors, silicone variance and the ability of "osseointegration" (a medical term derived through the direct structural and functional connection between bones and prosthesis) [3].

The dramatic uptake in research and innovation towards the development of highly-functional prosthetic limbs comes, as the issue retains prevalence on the global stage. It is estimated that one in 190 Americans lives with a partial or full loss of limb, with the number expected to more than double by 2050 [4]; this figure is liable in part due to: a large number of injuries through explosive devices in the multitude of ongoing global wars, extreme cases of peripheral arterial disease and congenital limb deficiencies [5]. The presented data signifies the prevalence of loss-of-limbs and highlights the absolute necessity of developing highly functional prosthetics for those with amputations and congenital issues.

Observing the economic and social imbalance above, inaccessibility to prosthesis provides a clear need for cost-reduction measures throughout the industry. One such method of doing so relies upon 'Fused Filament Fabrication (FFF) based 3D printing. FFF printing provides professionals with a means of developing multiple domains of medical devices that benefit from increased cost efficiency, customisation and enhanced productivity [6], including prosthetic limbs. Already, prosthetics have been developed with the inclusion of advanced functional technology such as integration with EEG for compound movements, while retaining a much lower price [7].

3D printing has been trialed and tested with companies such as Open Bionics who developed the Hero arm as a 3D printed alternative to contemporary prosthetics and now holds the title of the first medically approved prosthetic [8]. This shows a promising future for the technology furthered by the company winning the UAE AI & Robotics Award for Good of $1 million [9] reinforcing the need for low-cost prosthetics. E-nable is a global community of volunteers lending their 3D printers to print open sourced prosthetics for prospective patients [10]. Though these prostheses are very basic functions, relying on the wrist movement as a way to actuate fingers on the prosthetic hand, these low-cost prosthetics are ideal for children who quickly out grow their prosthesis [11].

Myoelectric prosthetics capture electric signals from the remaining muscles in a patient's residual limb where the signals are transferred to a controller within the prosthetic. The controller will then convert these electric signals into controls signals. Should the patient's muscle be unable to attach an EMG sensor, muscles elsewhere on the patient's body such as the chest can be used instead with muscle signals setup and defined to trigger movement on the prosthesis [12]. Most notable examples of myoelectric prosthetics are Open Bionics Hero arm, Ottobock BeBionic hand, and Mobius Bionics LUKE arm. These devices all employ the use of myoelectric technologies for movements within the prosthesis to execute movements for the user.

Classifying the data brought in through these sensors is done through windowing techniques, defining feature sets within the data that are unique to a specific movement of the patient's muscle. It is through these muscle sets that the patient can control multiple gestures by applying a demodulating algorithm to the input signal. The possibility that artificial intelligence could be used here in place of manual windowing techniques is tantalising, and shows great promise with the rapid development of AI techniques. This technology is seeing greater results than that of the R&D in EMG classification methods [13].

Pattern recognition is a machine learning process requiring predetermined knowledge recorded from patterns to inform a classifier to detect this information within input data. The concept of transfer learning is to recycle pre-trained models to be retrained on a new task saving the time and resources that would be required to train an entire network for the task. Instead, transfer learning focuses on the retraining of the final layer in the network relying on knowledge of the previous layers to inform the retraining process of the new task [14]. This method does have disadvantages in that negative transfer may occur. Negative transfer is a phenomenon whereby the accuracy of the network decreases after training, requiring adjustments to the training hyper-parameters [15]. Many pre-trained models are widely available for transfer learning deployments such as Inception, MobileNet or VGG [16-19].

## 2. Previous Work on Robot Arm Prostheitc

Understanding of existing technologies available on the market will help identify areas within the problem space that may help to provide requirements and functionality for the system. Three major prosthetic technologies were used as references behind our proposed research.

### 2.1. Open Bionics Hero Arm

Open bionics is a Bristol based company formed on the back of founder Joel Gibbard's enthusiasm for robotics and prosthetics [20]. The company aims to create affordable upper limb prosthetics and make them more accessible to those who require them, avoiding the traditional hooks and grippers but releasing a functional, bionic hand. This was realised in the form of the companies Hero arm, the first 3D printed prosthetic device medically approved in the UK. The company prides itself on the customisation of the Hero arm with detachable plates.

The Hero arm features what the company refers to as a dynamic socket; an area for users' residual limb to reside, along with a removable and ventilated liner for comfort. The arm is powered by an internal, rechargeable battery available in two different form factors (internal and external battery configurations). Improvements could be made to the arm as currently the arm operates using 'grip groups' which consists of a pair of available gestures per group. The arm is controlled through an open and close method, where the EMG sensors in the arm will pick up signals from the user's forearm. Gestures within a group can be switched by pressing and releasing the button on the back of the hand, though this method could be frustrating for a user depending on the movements needed [20].

### 2.2. BeBionic

Available in two sizes (original and small), the BeBionic hand [21] is touted by the company as the world's most lifelike bionic hand. The hand is moved through individual linear actuators for each finger; but requires the user's opposite hand to move the opposable thumb and the wrist into the desired position. Coupled with its construction from advanced materials, this makes the appendage capable of lifting 45kg. The BeBionic hand boasts 14 grip patterns and

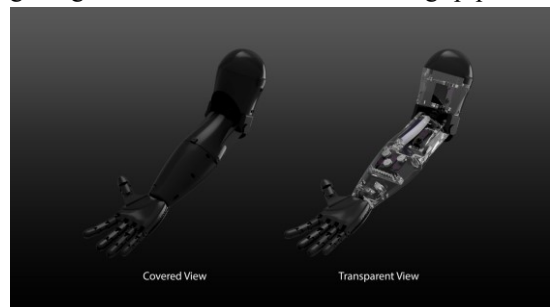

a) 3D Mesh Model of Prosthetic Arm

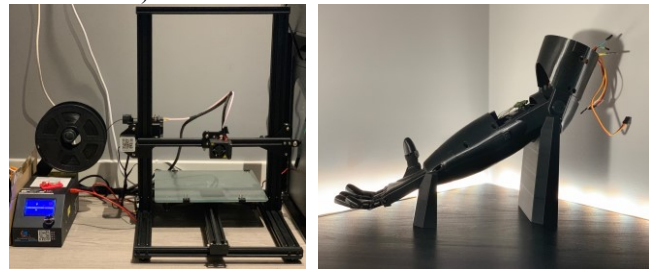

b) Our 3D printer    c) Product from our 3D printer
Figure 1. Our 3D Printed Robotic Arm Prosthetic.

is also the only prosthetic that provides multiple wrist connection options for detaching the hand from the forearm extension [21]. Although this prosthetic is comparably more advanced than the other two technologies reviewed, the cost of this prosthetic is considerable, quoted in a BBC article as saying the prosthesis costs around £30,000 [21].

### 2.3. Mobius Bionics Luke Arm

Mobius Bionics [22] is a medical device company developing advanced upper limb prosthetics for prospective patients. Their product, the LUKE arm is touted as the only commercially available prosthesis with a powered shoulder socket, containing ten powered joints. The arm also features six aptly named gestures/grip patterns pre-programmed into the appendage along with a range of available input device options. Of the three researched prosthetics this was the only one to use user input outside of EMG signals covering: inertial measurement unit (IMU) foot control, pressure transducer, pressure switch, rocker switch and linear transducer support.

## 3. Robot Arm Prosthetic Design

### 3.1 Hardware Development

Fig.1 shows our produced robot arm via a 3D printer. The analysis phase has outlined the core research requirements to enable us to begin the design phase. This section will highlight the different areas addressed during this phase, issues encountered and how the final implementation was reached.

#### 3.1.1. Hardware Architecture

This research is comprised of multiple hardware components that require communication between one another so as to function completely; an overview of this architecture can be seen in Fig.2. This map of the system highlights three distinct communication methods used within this research.

The Myo armband employs the Bluetooth protocol through a universal serial bus (USB) dongle connected to a personal computer (PC). The Myo is then paired to the computer using the 'Myo Connect App' provided with the Myo; this is required for the program to execute and will not display data without this step. A custom profile can be created using this software to calibrate the sensor more efficiently for each user.

Communication between the PC and the embedded Raspberry Pi with ARM processors is handled using the SSH protocol. Though this method may incur some latency depending on network speeds, it enables the use of the program across many machines without the need for adapters; the PC only uses the USB Type C connection, while one adapter is necessary for the Myo dongle this would mean a second would be needed to connect to the Raspberry Pi. Data sent over the SSH connection are encrypted meaning a secure connection can be made over any network the system connects to; the information sent is a Linux system level command that executes a python file corresponding to the gesture predicted by the network.

Finally, a wired connection directly from the Adafruit servo hat on the Raspberry Pi to the servos mounted within the robot arm model carries the pulse width modulation (PWM) commands to execute the movement predicted.

#### 3.1.2. 3D Printing Setup

Printing the model requires the use of software known as a slicer. The slicer is used to handle the 3D models to be printed by 'slicing' them into layers and producing G-code commands sent to the printer over a serial connection. While there are many slicers available, most of these come at an expense. Cura is a free, open-source slicer available under the LGPLv3 license developed by Ultimaker, a 3D printer manufacturing company. Though Ultimaker develops the slicer, settings can be customised to support all fused filament fabrication (FFF) printers.

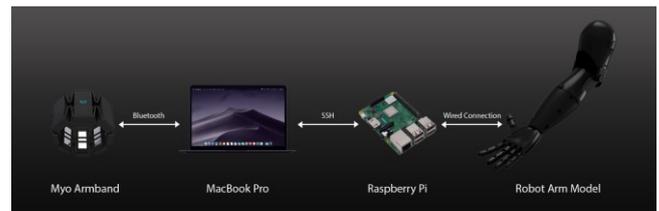

Figure 2: System Hardware Architecture

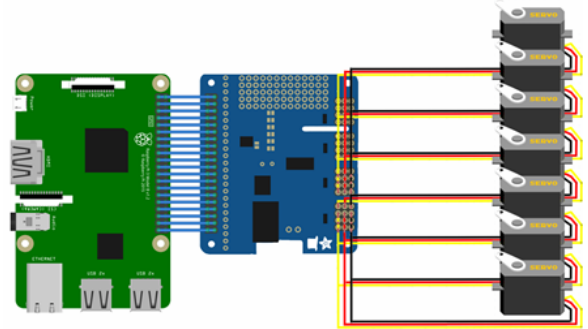

Figure 3: Robot Arm Wiring Diagram

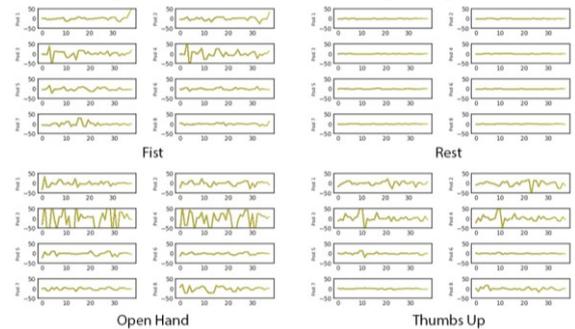

Figure 4 Sample sEMG Graph Data

### 3.1.3. SSH Communication

Communication between the host computer and the embedded Raspberry Pi was needed, and though this could have been solved using a serial connection through USB cable to the author's computer, the Raspberry Pi requires adapters and further pins to be used to allow this. As the Adafruit PWM/Servo hat sits directly over the GPIO pins on the Raspberry Pi, this would have proved challenging. SSH allowed for wireless communication, meaning that should this research continues in the future, a wireless communication method has already been implemented. SSH was implemented into the system using the Paramiko package extension and works on any connected network, employing encryption techniques to transferred data to ensure privacy and anonymity; this is important if the system was ever connected to public wi-fi, especially if this was a marketed product due to GDPR issues regarding the user's EMG data, as shown in Fig.4.

The decision was made to continue with the SSH protocol due to its future proofing of the work, should it be taken further and no extra hardware purchases necessary; the PC employed only has USB-C connections and would require a USB-A to USB-C adapter to use the serial method, as well as a UART-USB adapter for the Raspberry Pi to transfer data. Negatives to this method include network speeds and signal strength contribute to the speed in which data is transferred between devices, and as this system must operate as close to real-time as possible this could become a problem. However, testing of the system with the SSH protocol showed very fast response speed.

### 3.1.4. Movement

Movement of the arm is handled via seven servos housed in various places throughout the arm. This research is aimed at using transfer learning to predict hand gestures, and as so, two of these servos are redundant. To trigger the movement of the fingers in the arm, multiple python files are stored on the embedded Raspberry Pi written using CircuitPython; an extended version of Python used for programming with microcontrollers. These files are executed through the Predictor class when a predicted gesture is thrown by the model, the returned data is passed through a series of if statements where the corresponding gesture label will create a new connection object to connect to the Raspberry Pi over SSH and pass a system level command to the Linux terminal, executing the python file and triggering the movement.

## 3.2. Software Development
### 3.2.1. Training

Training of the system was an iterative process, with multiple training sessions attempted and compared to the previous efforts so as to find the optimal training configuration. TensorFlow's retrain.py file was used to train the model over a total of six training sessions applied to the system, starting with TensorFlow's default setting of 4000 epochs. TensorBoard generates data required to compare each training session precompiled into graphs where accuracy and cross entropy of each model can be analysed to highlight potential further improvement. The fourth training attempt of the system was performed on 300,000 epochs where accuracy and cross-entropy began to plateau, as shown in Fig.5. Supplementary testing was performed with different configurations of batch size and learning rates to optimise the system further.

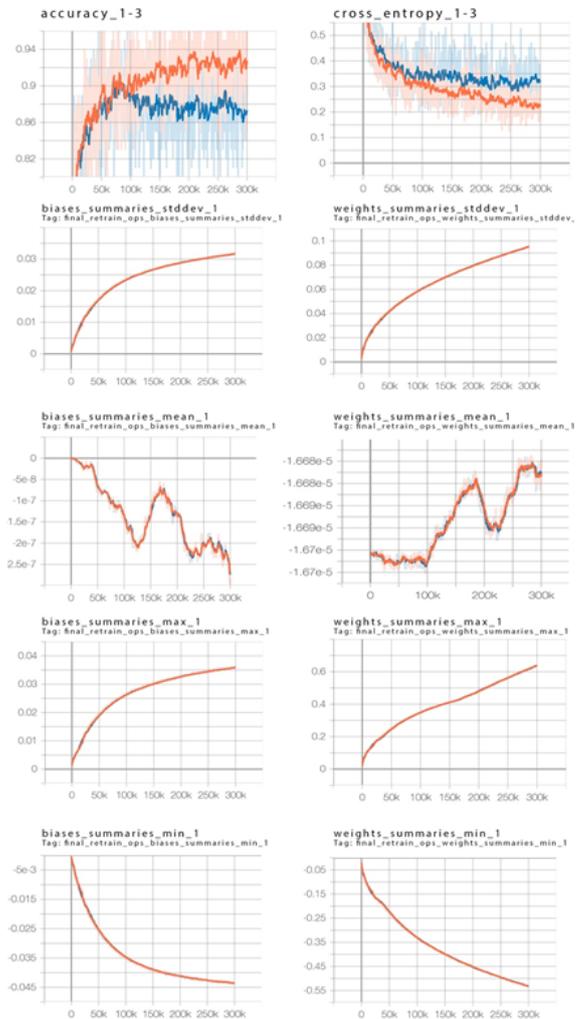

Figure 5. Training Process

Table I Accuracy Rates and Cross-Entropy Results

|  | Accuracy | Cross-Entropy |
|---|---|---|
| 4,000 | 0.89 | 0.35 |
| 25,000 | 0.93 | 0.2 |
| 100,000 | 0.95 | 0.15 |
| 300,000 | 0.98 | 0.075 |
| 300,000 + LR Changes | 0.93 | 0.22 |
| 300,000 + LR and BS | 0.93 | 0.22 |

3.2.2. Prediction

Using the TensorFlow framework as the basis for the AI implementation for the classification of EMG images, the project initially relied on the use of the label_image.py file to make predictions on incoming data in real-time. This produced a problem as the label_image.py file contains a call to load the trained model every time a prediction is made. As this affected the system performance and meant the real-time prediction requirement could not be met, this file was modified into what became the predictor class.

## 4. Experimental Results

### 4.1. Testing Strategy

Testing of the system is an important part of development regardless of the chosen methodology. It allows for errors to be discovered, optimisation and can reveal potential flaws within a system use case scenario's as well as detail user expectancies of the system.

We have adopted the agile methodology for this research and as such, testing is required during each iteration of development. These tests produced no useable data but did provide valuable insight into future iterations and direction for the software.

### 4.2. Training and Performance Testing

Tests devised focused on two areas, training and test performances. Results for the training of the system provided through TensorBoard and served as a visual representation of the model's accuracy based on pre-provided training/testing data. The test set is used to test model efficiency.

Strategies for testing and training of the system were designed to start with TensorFlow's default training settings; from here results could be analysed with TensorBoard to give us a baseline for the system. The experiments proceed with exponential increases in epoch values, leading to tests performed within a range of 4000 – 300,000 epochs where the accuracy of the system began to plateau. From here, multiple hyper-parameters were adjusted to provide further optimisation of the model at 300,000 epochs; these parameters included training image batch size and learning rates. The training performance is shown in Figure 5.

### 4.3. Test Results

The results of the performance testing were captured using TensorBoard, a local-HTML page included with the TensorFlow framework. Data obtained during the training of the model are presented in graph form, showing accuracy and cross-entropy details as well as further insights such as standard deviation. The results have also been collated into a table shown in Table I.

The results were promising as adjustments to the 300,000 epoch models. Further testing would be required to see if training image batch size has any effect on accuracy and cross-entropy values of the model. The results of these graphs landed almost identically to the 25,000 epoch model though it took 6 times the amount of time required to train.

Although the 300,000 epoch model plateaued, its accuracy value suspended around 98% which is a very acceptable score for the system based on its accuracy requirement.

## 5. Conclusion

This research was designed as a proof of concept to show how transfer learning could be used to accurately predict movements/gestures from incoming EMG signals. Specifically, it employs transfer learning to accurately classify incoming EMG data presented in a graph form. The input images were used to retrain the final layer of Google's Inception v3 CNN and allowed for accurate prediction of incoming EMG data within 98% accuracy. There are however interesting results shown during retraining with hyper parameter fine-tuning, where the accuracy rates decrease during the experiments in comparison to those with the default settings. This kind of optimisation is worth investigating further as the model began to plateau at around 300,000 epochs. Other strategies and mechanisms will be explored in future directions to address this issue [32-45].